\documentclass[lettersize,journal]{IEEEtran}
\usepackage{amsmath,amsfonts}
\usepackage{algorithm2e}
\usepackage{array}
\usepackage[caption=false,font=normalsize,labelfont=sf,textfont=sf]{subfig}
\usepackage{textcomp}
\usepackage{stfloats}
\usepackage{url}
\usepackage{verbatim}
\usepackage{graphicx}
\usepackage{cite}

\usepackage{makecell}

\usepackage{graphicx} 
\usepackage{amsmath}
\usepackage{amsmath}
\usepackage{amsmath, amssymb}
\usepackage{xcolor}	

\usepackage{algpseudocode}
\usepackage{graphics}
\usepackage{amsmath, amsthm}

\usepackage[caption=false,font=normalsize,labelfont=sf,textfont=sf]{subfig}

\newcounter{assumption}

\newcounter{theorm} 

\begin{document}
	
	\title{Fluid Antenna Systems under Channel Uncertainty and Hardware Impairments: Trends, Challenges, and Future Research Directions}

	\author{IEEE Publication Technology,~\IEEEmembership{Staff,~IEEE,}
	}
	
	

	\author{Saeid Pakravan, Mohsen Ahmadzadeh, Ming Zeng, Wessam Ajib, Ji Wang, Xingwang Li

\thanks{S. Pakravan and W. Ajib are with the Department of Computer Science, Université du Québec à Montréal (UQAM), Montreal, QC, Canada. email: pakravan.saeid@uqam.ca; ajib.wessam@uqam.ca.}

\thanks{M. Ahmadzadeh is with the Department of Electric and Computer Engineering, Ferdowsi University, Mashhad, Iran. email: m.ahmadzadehbolghan@mail.um.ac.ir.}

\thanks{M. Zeng is with the Department of Electric and Computer Engineering, Laval University, Quebec City, QC, Canada. email: ming.zeng@gel.ulaval.ca.}

\thanks{J. Wang is with the Department of Electronics and Information Engineering, College of Physical Science and Technology, Central China Normal University, Wuhan 430079, China. email: jiwang@ccnu.edu.cn.}

\thanks{X. Li is with the School of Physics and Electronic Information
Engineering, Henan Polytechnic University, Jiaozuo 454000, China. email:
lixingwang@hpu.edu.cn.}

	}

	\maketitle

	\begin{abstract} Fluid antenna systems (FAS) have recently emerged as a promising paradigm for achieving spatially reconfigurable, compact, and energy-efficient wireless communications in beyond fifth-generation (B5G) and sixth-generation (6G) networks. By dynamically repositioning a liquid-based radiating element within a confined physical structure, FAS can exploit spatial diversity without relying on multiple fixed antenna elements. This spatial mobility provides a new degree of freedom for mitigating channel fading and interference, while maintaining low hardware complexity and power consumption. However, the performance of FAS in realistic deployments is strongly affected by channel uncertainty, hardware nonidealities, and mechanical constraints, all of which can substantially deviate from idealized analytical assumptions. This paper presents a comprehensive survey of the operation and design of FAS under such practical considerations. Key aspects include the characterization of spatio-temporal channel uncertainty, analysis of hardware and mechanical impairments such as RF nonlinearity, port coupling, and fluid response delay, as well as the exploration of robust design and learning-based control strategies to enhance system reliability. Finally, open research directions are identified, aiming to guide future developments toward robust, adaptive, and cross-domain FAS design for next-generation wireless networks.

\end{abstract}


\section{Introduction}

The emergence of sixth-generation (6G) wireless communication systems is widely recognized as a paradigm shift toward intelligent, immersive, and ubiquitous connectivity, enabling transformative applications such as autonomous transportation, digital twins, extended reality, and massive-scale Internet of Things (IoT) networks \cite{8782879, 9854866}. To support these applications, 6G networks are expected to achieve unprecedented performance targets, including data rates approaching the terabit-per-second regime, end-to-end latencies on the order of sub-milliseconds, ultra-high reliability, and native integration of communication, sensing, and computing functionalities \cite{8869705, 10812728}. Achieving these ambitious and often conflicting objectives requires disruptive advances across multiple layers of the wireless system, particularly in antenna and transceiver design, which form the fundamental interface between the electromagnetic environment and digital signal processing.

Over the past decade, multiple-input multiple-output (MIMO) systems and reconfigurable intelligent surfaces (RISs) have played pivotal roles in improving spectral efficiency, coverage, and energy efficiency in wireless networks \cite{8241348, 10480333,  9360709, 10363658, 8741198}. MIMO architectures exploit spatial multiplexing and diversity using multiple radiating elements and radio-frequency (RF) chains, whereas RISs enable partial control over the wireless propagation environment through programmable passive elements. Despite their notable success, both paradigms exhibit inherent structural limitations. Conventional MIMO systems suffer from high hardware cost, power consumption, and calibration complexity due to the reliance on multiple RF chains and fixed antenna geometries \cite{7389996, Bjornson2017MassiveMIMO}. RIS-based solutions, on the other hand, are constrained by static physical layouts, limited spatial adaptability, and predominantly passive operation \cite{Wei2023RISModelBasedModelFree, 10596064, 10080950}. These limitations restrict scalability and responsiveness in highly dynamic, dense, and space-constrained 6G scenarios, motivating the exploration of alternative antenna architectures that provide continuous spatial flexibility with reduced hardware overhead.

In this context, fluid antenna systems (FAS) have recently emerged as a promising and unconventional antenna paradigm that introduces dynamic reconfigurability at the antenna level \cite{9264694, 11302793}. Unlike conventional fixed-geometry antenna arrays, FAS exploit the ability of a single or a small number of radiating elements to dynamically reposition within a confined physical region, typically enabled by conductive fluids or mechanically tunable structures. By selecting an optimal antenna position (or ``port'') in response to instantaneous channel conditions, FAS can effectively harness spatial diversity without requiring multiple RF chains or complex array processing \cite{9131873, 10146274, 9715064}. This unique capability allows FAS to capitalize on small-scale fading variations, mitigate interference, and enhance link reliability in a cost- and energy-efficient manner, making them particularly attractive for spectrum- and energy-constrained wireless systems \cite{11247926, 10146286}.

Despite their conceptual appeal and demonstrated potential, the performance of FAS in realistic wireless environments remains insufficiently explored. Most existing studies rely on idealized assumptions, including perfect channel state information (CSI), linear and distortion-free RF hardware, instantaneous antenna repositioning, and negligible mechanical or fluidic dynamics \cite{Psomas2023FluidAntenna, 10807122}. In practical deployments, however, channel uncertainty arising from estimation errors, feedback latency, spatial correlation mismatch, and temporal channel variations can lead to suboptimal port selection and significant performance degradation. Furthermore, hardware impairments such as in-phase/quadrature (I/Q) imbalance, nonlinear amplification, RF noise, and mutual coupling introduce additional deviations from ideal behavior, particularly in compact, low-cost, or highly miniaturized FAS implementations \cite{7106472, 10114591, 7400949}. Fluid-based and mechanically tunable antenna architectures also suffer from response delays, thermal effects, viscosity-induced damping, and mechanical deformation, which jointly affect antenna impedance, radiation patterns, and spatial mobility \cite{Wang2025FluidAntenna, Khan2017ReconfigurablePatch, 7086418}. These effects introduce a nontrivial coupling between the electromagnetic, mechanical, and control domains, substantially complicating analytical modeling and performance prediction. At present, the absence of unified and physically consistent frameworks that jointly account for statistical channel uncertainty, hardware nonidealities, and mechanical constraints constitutes a critical gap in the FAS literature.

Although several recent survey and tutorial works have examined FAS from perspectives such as fundamental principles, spatial diversity, system-level integration, and emerging 6G applications \cite{11302793, 11247926, 10753482, 11175437, Shah2024FluidAntennaMA}, these studies largely rely on idealized assumptions or address practical nonidealities only in a fragmented or qualitative manner. In particular, a unified and systematic treatment of channel uncertainty, hardware impairments, and mechanical constraints and their coupled impact on FAS modeling, optimization, and performance remains missing. As summarized in Table~\ref{tab:FAS_survey_general}, existing surveys primarily focus on conceptual, architectural, or application-level aspects of FAS, while robust design methodologies for antenna position selection, adaptive beamforming, and resource allocation under imperfect, time-varying, and hardware-constrained conditions are still insufficiently explored. Addressing these gaps is essential for transitioning FAS from conceptual and experimental prototypes toward reliable, scalable, and deployable solutions in future 6G and IoT-oriented wireless networks.

Motivated by these observations, this paper presents a comprehensive survey of FAS with a particular emphasis on channel uncertainty and hardware impairments, aiming to provide uncertainty-aware insights and design guidelines for practical FAS deployment.

\begin{table*}[t]
\centering
\caption{Summary of Existing Survey and Tutorial Papers on FAS}
\label{tab:FAS_survey_general}
\renewcommand{\arraystretch}{1.2}
\setlength{\tabcolsep}{6pt}
\begin{tabular}{|p{0.6cm}|p{2.6cm}|p{6.2cm}|p{6.8cm}|}
\hline
\textbf{Ref.} & \textbf{Theme} & \textbf{Limitations} & \textbf{Main Contribution} \\
\hline

\cite{11302793} 
& FAS principles and 6G applications 
& Primarily vision- and application-oriented; channel uncertainty and hardware nonidealities are not systematically analyzed. 
& Surveys fundamental principles of FAS, representative applications, and high-level research directions toward 6G systems. \\ \hline

\cite{11247926} 
& System-level perspective on FAS 
& Largely relies on ideal or quasi-ideal assumptions; robustness under imperfect CSI and hardware impairments is not a central focus. 
& Provides a comprehensive system-level overview positioning FAS as a reconfigurable wireless communication paradigm. \\ \hline

\cite{10753482} 
& Comprehensive survey of FAS
& Broad coverage with limited emphasis on unified modeling of channel uncertainty and hardware impairments. 
& Reviews FAS from fundamentals to networking perspectives, highlighting integration challenges and opportunities. \\ \hline

\cite{11175437} 
& Fundamentals and networking perspectives 
& Focuses on baseline models and concepts; practical nonidealities and robustness aspects are only partially addressed. 
& Introduces core FAS concepts, port selection mechanisms, and fundamental performance insights. \\ \hline

\cite{Shah2024FluidAntennaMA} 
& Survey on fluid antenna multiple access 
& Scope limited to multiple access scenarios; general FAS uncertainty and hardware/mechanical constraints are outside its focus. 
& Surveys FAS-based multiple access techniques and diversity gains in compact multiuser systems. \\ \hline

\textbf{This work} 
& {Uncertainty-aware FAS design} 
& {---} 
& {Provides a unified survey of FAS under channel uncertainty, hardware impairments, and mechanical constraints, emphasizing robust and adaptive system design.} \\ \hline

\end{tabular}
\end{table*}

\begin{itemize}
    \item Categorize existing FAS channel models and identify key sources of uncertainty, including correlation mismatch, estimation errors, and temporal variability;

    \item Investigate the physical and electrical origins of hardware impairments, including RF noise, nonlinear distortion, coupling effects, and mechanical imperfections;

    \item Examine robust and adaptive control strategies that enhance system reliability in uncertain and imperfect conditions;
    
    \item Identify open research challenges and future directions for the design of uncertainty-aware and physically consistent FAS frameworks.

\end{itemize}

By consolidating and analyzing current knowledge, this review highlights critical research gaps and identifies open challenges, guiding future work toward practical, uncertainty-aware FAS designs capable of reliable operation in dynamic and imperfect wireless environments.

The remainder of this paper is organized as follows. Section II presents an overview of FAS architectures and their fundamental operational principles. Section III reviews existing channel modeling approaches, highlights the primary sources of uncertainty, and examines hardware impairments and mechanical constraints that influence FAS performance. Section IV discusses current strategies for robust optimization and adaptive control under uncertain environments. Section V outlines open research challenges and presents future research directions. Finally, Section VI concludes the paper.

\section{Fundamentals of FAS}

\subsection{Concept and Operating Principle}

\begin{figure}[]
    \centering
    \includegraphics[width=7.5cm, height=4.2cm]{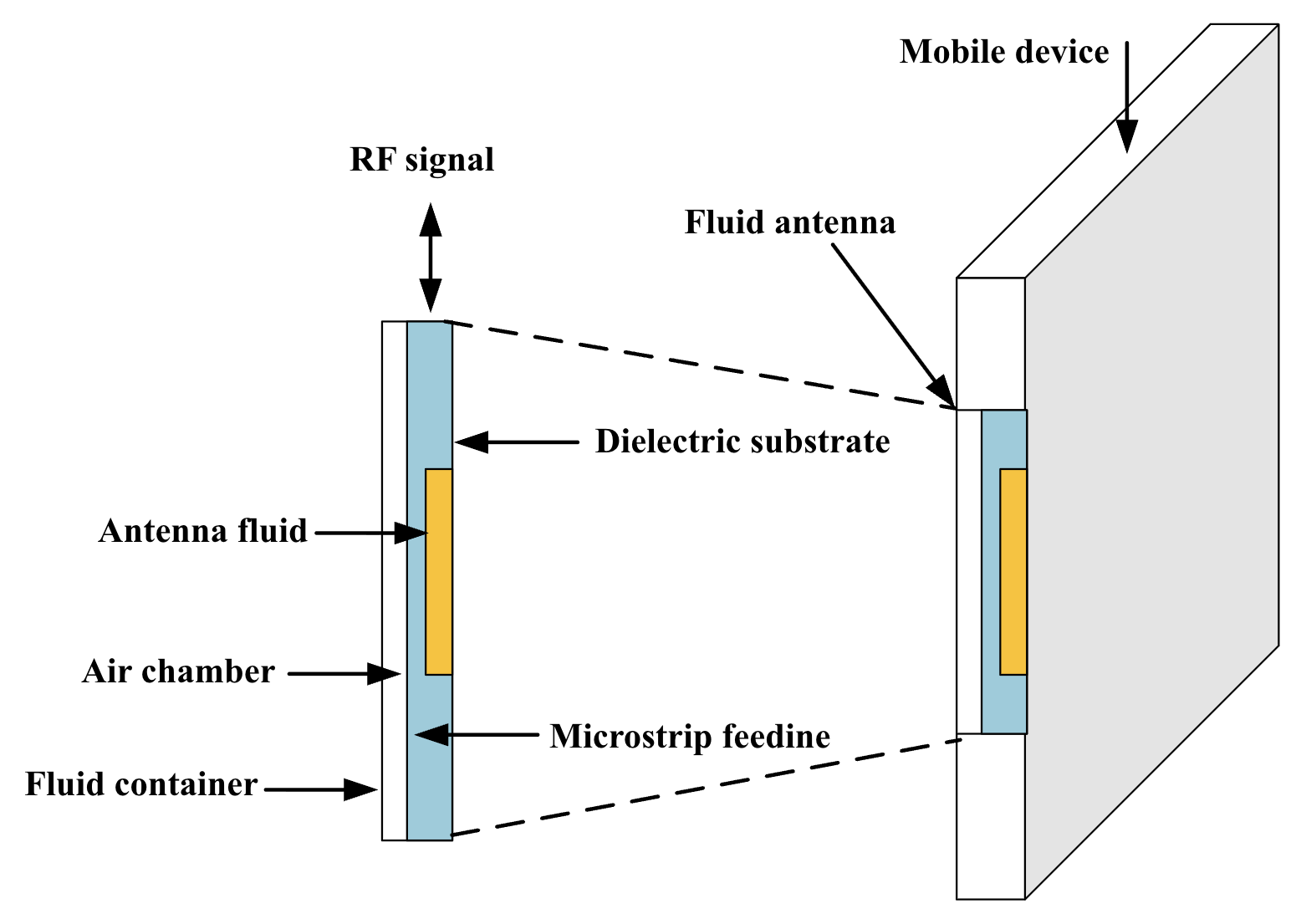} 
    \caption{Illustration of an FAS concept implemented on a mobile device.}
    \label{fig:FAS_concept}
\end{figure}

FAS constitute a novel class of reconfigurable antenna architectures that exploit spatial fluidity rather than static geometry \cite{9264694, 11302793, 9131873, 10146274, 9715064, 11247926, 10146286}. Unlike traditional antenna arrays where radiating elements are fixed in position, FAS allows its radiating structure-typically realized through liquid metal, conductive fluid, or electrochemically tunable materials-to dynamically change position, shape, or electrical length within a predefined region known as the fluid region, as illustrated in Fig.~\ref{fig:FAS_concept}. By repositioning a single radiating element across multiple candidate ports, the FAS can select the location that provides the most favorable instantaneous channel gain \cite{9715064, 10375698, 10354003}.

This dynamic spatial mobility allows FAS to transform spatial diversity into a controllable temporal mechanism. Since small-scale fading varies across space even within compact regions, the antenna can rapidly explore these spatial variations without requiring multiple RF chains or complex beamforming circuitry. Instead of treating fading as an impairment, FAS convert it into an exploitable resource for diversity enhancement.

Let $h^{(t)}[x]$ denote the complex channel coefficient measured at position $x$ and time $t$. The optimal antenna position $x^{\star}$ is then selected according to
\begin{equation}
    x^{\star} = \arg\max_{x \in \mathcal{X}} \big| h^{(t)}[x] \big|,
\end{equation}
where $\mathcal{X}$ represents the set of feasible fluid-region positions. 
Depending on the operating constraints, this port-selection process may be performed instantaneously using real-time CSI or statistically based on 
long-term channel characteristics. The ability to physically adapt the antenna’s location in this manner provides a compact and energy-efficient mechanism for achieving substantial spatial diversity gains.

\subsection{Architecture and Implementation Approaches}

A variety of hardware platforms have been developed to realize controllable fluidic or pseudo-fluidic antenna mobility. These approaches differ in reconfiguration speed, integration feasibility, mechanical reliability, and robustness to channel and hardware uncertainty. The most common approaches include:

\begin{itemize}
    \item \textbf{Electrohydrodynamic and Electrocapillary Designs:} In these designs, a conductive liquid metal (e.g., eutectic gallium–indium, EGaIn) is manipulated within microchannels by applying an electric potential or pressure gradient. The position of the liquid determines the active port, allowing smooth reconfiguration across several millimeters with high precision \cite{9264694, 10753482}.
    
    \item \textbf{Microfluidic and Dielectric Channel Structures:} Microfluidic-based FAS architectures embed the antenna in soft polymer substrates, where the liquid conductor can flow to predefined cavities or ports \cite{Tong2025FASHardware, 10740058}. These structures are particularly attractive for integration with flexible or wearable electronics.
    
    \item \textbf{Electronically Switched FAS:} Instead of physical liquid flow, this configuration employs electronically controlled switching among multiple miniaturized radiators that emulate fluidic behavior \cite{Zhang2024PixelFAS, Wang2025FluidAntenna}. It offers faster reconfiguration speed at the cost of additional circuitry.
    
    \item \textbf{Hybrid Mechanically Tunable Systems:} In certain prototypes, mechanical actuators or thermal expansion elements adjust the physical position or geometry of the antenna, achieving reconfigurability with minimal electrical complexity \cite{Qian2021LiquidMetalMetasurface, 9131873}.
\end{itemize}

Each architecture involves different trade-offs between reconfiguration speed, implementation complexity, energy efficiency, and resilience to mechanical or thermal perturbations. A concise comparison of the principal FAS architectures is summarized in Table~\ref{tab:FAS_architecture}. The selection of a specific implementation strategy directly affects the level of uncertainty and potential hardware impairments that must be considered in practical system design.

\begin{table*}[t]
\centering
\caption{Comparison of Major FAS Architectures}
\label{tab:FAS_architecture}
\renewcommand{\arraystretch}{1.15}
\setlength{\tabcolsep}{5pt}
\begin{tabular}{|p{1.2cm}|p{2.7cm}|p{5.05cm}|p{2.2cm}|p{2.2cm}|p{2.3cm}|}
\hline
\textbf{Ref.}
& \textbf{Architecture Type}
& \textbf{Core Mechanism}
& \textbf{Reconfiguration Speed}
& \textbf{Energy Efficiency}
& \textbf{Design Complexity} \\ \hline

\cite{9264694,10753482}
& Electrohydrodynamic
& Liquid flow manipulated by electric field
& Medium
& High
& Moderate \\ \hline

\cite{Tong2025FASHardware,10740058}
& Microfluidic
& Conductive liquid within a polymer substrate
& Slow
& Very High
& Low \\ \hline

\cite{Zhang2024PixelFAS,Wang2025FluidAntenna}
& Electronically Switched
& Rapid switching among discrete radiators
& Very High
& Moderate
& High \\ \hline

\cite{Qian2021LiquidMetalMetasurface,9131873}
& Mechanical / Hybrid
& Physical displacement or thermal actuation
& Low
& High
& Moderate \\ \hline

\end{tabular}
\end{table*}

\subsection{Performance Advantages and Design Features}

The distinct operation of FAS provides several notable advantages over conventional antenna systems:

\begin{itemize}
    \item \textbf{Enhanced Diversity Gain:} FAS achieve spatial diversity comparable to multi-antenna arrays while employing only a single RF chain, significantly reducing cost, energy consumption, and form factor \cite{10130117, 10753482, 11175437}. This makes FAS particularly attractive for IoT devices, wearables, and small mobile terminals.
    
    \item \textbf{Interference Mitigation:} By repositioning the radiating element to locations with stronger channel gains or lower interference, FAS effectively avoid deep fades and interference hotspots, enhancing link robustness in dense or dynamically obstructed environments \cite{Tong2025FASHardware, 11258084}.
    
    \item \textbf{Reduced Implementation Complexity:} Since FAS can emulate array behavior without multiple RF front-ends, it significantly reduces circuit complexity and energy consumption, making it suitable for IoT, vehicular, and mobile applications \cite{Zhang2025CSIFluidAntenna, 11206363}.
    
    \item \textbf{High Adaptability to Environmental Dynamics:} FAS can respond to temporal channel variations caused by user mobility or environmental changes, ensuring consistent performance without centralized coordination \cite{10299674, Zhang2023FastPortFAS}.

    \item \textbf{Scalability and Integration Flexibility:} FAS enable compact integration in flexible, wearable, or miniaturized platforms, while hybrid fluidic-electronic designs enhance adaptability. This versatility supports scalable deployment in dense or heterogeneous networks and allows seamless integration with advanced physical-layer techniques such as non-orthogonal multiple access (NOMA) and RIS-assisted communications \cite{10318134, 10423153,  10794591}.

\end{itemize}

These characteristics position FAS as an efficient, flexible, and scalable technology capable of complementing or even replacing conventional diversity techniques in future wireless systems.

\section{Channel Uncertainty and Hardware Impairments in FAS}

FAS exploit the dynamic reconfigurability of conductive fluids to achieve enhanced spatial diversity and spectral efficiency. However, the practical performance of FAS is fundamentally limited by two interrelated factors: channel uncertainty and hardware impairments \cite{10751774, Zhu2025FluidAntennaGeo, 10375559, 11224643, Li2025HardwareImpairedFAS, Wang2025FluidAntenna, Irshad2025MovableAntennaMIMO, Tlebaldiyeva2024FASNOMA,  Pakravan2025FluidAntennaNOMA, Hong2025GeometricShapingFAS, Pakravan2024OTAComputationFAS, 10807122}. Accurate characterization and mitigation of these factors are critical for robust port selection, reliable communication, and system optimization. This section provides a comprehensive overview of the sources, modeling frameworks, impact, and mitigation strategies for both channel uncertainty and hardware imperfections.

\subsection{Channel Uncertainty}

\subsubsection{Source of Channel Uncertainty}
Channel uncertainty arises due to imperfect knowledge of the wireless propagation environment. In FAS, the position of the fluidic antenna directly influences the received signal strength and interference profile, making accurate CSI essential.

\textbf{Estimation Errors:}  
In most FAS implementations, the receiver or access point estimates the instantaneous channel coefficients at different antenna positions. Due to noise, limited pilot length, and hardware constraints, these estimates are inherently imperfect. The estimated channel can be represented as 
\begin{equation}
    \hat{h}^{(t)}[x]=h^{(t)}[x]+\varepsilon^{(t)}[x],
\end{equation}
 where $h^{(t)}[x]$ is the true channel coefficient and $\varepsilon^{(t)}[x]$ denotes the estimation error, often modeled as a complex Gaussian random variable with variance proportional to the signal-to-noise (SNR) \cite{10751774, 11224420, 10906511}. 

\textbf{Temporal Variability:}  
The fluid movement and user mobility introduce temporal variations that cause the channel to decorrelate over time \cite{10207934}. When the FAS operates with non-negligible fluid response delay or when channel estimation and port selection occur at different time instances, outdated CSI can lead to performance degradation. The channel time evolution can be expressed using a first-order autoregressive model as \cite{1715541, marzetta2016massive, 9377710} 
\begin{equation}
    h^{(t+\Delta t)}[x]=\rho_t h^{(t)}[x]+\sqrt{1 - \rho_t^2}\, z^{(t)}[x],
\end{equation}
where $\rho_t$ denotes the temporal correlation coefficient and $z^{(t)}[x]$ represents a zero-mean complex Gaussian innovation term.

\textbf{Spatial Correlation Mismatch:}  
The underlying assumption of spatial channel correlation within the fluid region is essential for port diversity analysis. However, practical channels may deviate from theoretical correlation models (e.g., Jakes or exponential models) due to environmental irregularities or non-isotropic scattering. This correlation mismatch reduces the expected diversity order and modifies the spatial fading distribution, making it difficult to predict FAS performance accurately \cite{Martinez2024CorrelationFAS, 10623405, Wong2022SpatialCorrelationFAS}.

\textbf{Feedback and Quantization Errors:}  
When FAS operates in a closed-loop configuration, CSI or port index information must be fed back to the transmitter. Limited feedback bandwidth and quantization cause additional errors that compound estimation and temporal uncertainties, further distorting port selection decisions \cite{11175437, Wang2025FluidAntenna, Faddoul2025ReconfigurableArrays}.

\subsubsection{Modeling Frameworks of Channel Uncertainty}

To analyze and mitigate the impact of channel uncertainty, various modeling frameworks have been adopted in the literature \cite{10092780, 10751774, Li2025HardwareImpairedFAS, 11224643, 10207934}.

\textbf{Stochastic Error Models:}  
In stochastic formulations, the channel estimation error $\varepsilon^{(t)}[x]$ is treated as a random variable with known distribution (commonly Gaussian) \cite{10751774, Pakravan2024OTAComputationFAS}. This approach facilitates probabilistic performance analysis, such as the derivation of average error probability or outage probability under random channel perturbations.

\textbf{Bounded or Worst-Case Models:}  
In robust optimization, the uncertainty is modeled as belonging to a deterministic set, i.e., $\mathcal{H} = \{ h : \| h - \hat{h} \| \leq \delta \}$, where $\delta$ represents the uncertainty radius. The design objective is to optimize the system performance under the worst-case realization within this set, ensuring guaranteed robustness against estimation or feedback errors \cite{Li2025HardwareImpairedFAS, 10092780, Wu2025FASAAVSecurity}.

\textbf{Hybrid Stochastic–Deterministic Models:}  
Recently, hybrid models that combine statistical and bounded uncertainty have been proposed to capture both random noise and systematic modeling bias \cite{11224643, 8472907, 7106472, 11175437}. Such models are particularly relevant for FAS, where uncertainty stems not only from measurement errors but also from imperfect knowledge of the physical relationship between antenna position and channel response.

\subsubsection{Impact of Channel Uncertainty on System Performance}

Channel uncertainty directly affects the ability of FAS to identify the optimal antenna location. Even small errors in channel estimation or spatial correlation may lead to incorrect port selection, resulting in a substantial loss of received SNR. For instance, assuming an ideal selection based on perfect CSI yields an SNR of $\gamma_{\text{ideal}} = \max_{x \in X} \frac{P |h^{(t)}[x]|^2}{\sigma^2}$, while in the presence of imperfect CSI, the realized SNR becomes $\gamma_{\text{imperfect}} = \frac{P |h^{(t)}[x^{*}]|^2}{\sigma^2}$, where $x^{*}$ denotes the port chosen based on the estimated $\hat{h}^{(t)}[x]$. The relative loss can be quantified by
$\Delta \gamma = 10 \log_{10} \left( \frac{\gamma_{\text{imperfect}}}{\gamma_{\text{ideal}}} \right)$,
which highlights how sensitive FAS performance is to even modest estimation errors or feedback delay \cite{Zhang2023FastPortFAS, 1284943}.

Moreover, uncertainty alters the statistical diversity gain that FAS can achieve. When the correlation mismatch increases, the effective number of independent spatial samples decreases, limiting the achievable diversity order. Consequently, analytical models developed under perfect CSI assumptions tend to overestimate the actual reliability and capacity of FAS systems.

\subsubsection{Strategies for Uncertainty Mitigation}

Mitigating channel uncertainty is essential for unleashing the full potential of FAS. Because FAS rely on ultra-fast and fine-grained spatial reconfiguration, even small CSI mismatches, outdated estimates, or modeling errors can significantly degrade port selection, increase error rates, and reduce the achievable diversity gains. Several mitigation techniques have been investigated in adaptive arrays, reconfigurable antennas, and location-based beamforming, many of which can be directly extended or tailored to the distinct characteristics of FAS.

\begin{itemize}

\item \textbf{Adaptive Channel Estimation and Tracking:}  
Real-time tracking of small-scale channel fluctuations is crucial for FAS, where the optimal port may change rapidly with user mobility or environmental dynamics \cite{Zhang2023FastPortFAS}. Methods such as recursive least squares, Kalman filtering, particle filtering, and lightweight neural predictors have been used to reduce the impact of noisy or delayed estimates. These techniques enable continuous refinement of CSI, compensate for feedback delay, and improve the accuracy of spatial gain predictions across fluid antenna positions.

\item \textbf{Robust Port Selection and Optimization:}  
Traditional port selection assumes perfect CSI, which is unrealistic in practical FAS. Robust optimization frameworks such as worst-case optimization, chance-constrained formulations, and Bayesian port ranking explicitly incorporate uncertainty sets or estimation error distributions \cite{Hu2025UplinkFASMIMO, Tong2025FASHardware, 11192669}. By designing port-selection rules that remain reliable even under CSI distortions, these approaches help maintain system performance and prevent catastrophic selection errors due to noise, quantization, or outdated feedback \cite{ 10092780, Wu2025FASAAVSecurity}.

\item \textbf{Statistical Learning and Model Calibration:} 
Misalignment between assumed and actual spatial correlation characteristics can lead to suboptimal mobility patterns in FAS \cite{11175437, 10751774, Pakravan2024OTAComputationFAS, 10146274}. Data-driven calibration, online learning, Gaussian-process regression, and spatial kernel estimation can be used to infer the true correlation structure and update model parameters over time \cite{8741198, 1306481, 8680025, 8786074}. Such calibration reduces model mismatch, improves prediction accuracy for candidate port gains, and adapts the FAS sensing strategy to environmental non-stationarities.

\item \textbf{Diversity-Oriented Combining under Uncertainty:} 
When CSI quality is low or rapidly varying, relying on a single port selection may be fragile. Techniques such as maximum-ratio combining, equal-gain combining, or selection combining across multiple fluid positions can enhance robustness \cite{1284943, 10375698}. These schemes exploit the inherent spatial sampling capability of FAS, offering resilience against estimation errors by averaging or combining multiple imperfect spatial observations rather than depending on a single possibly misestimated port.

\end{itemize}

\begin{figure*}[t!]
\centering
\includegraphics
[width=18cm, height=10.7cm]
{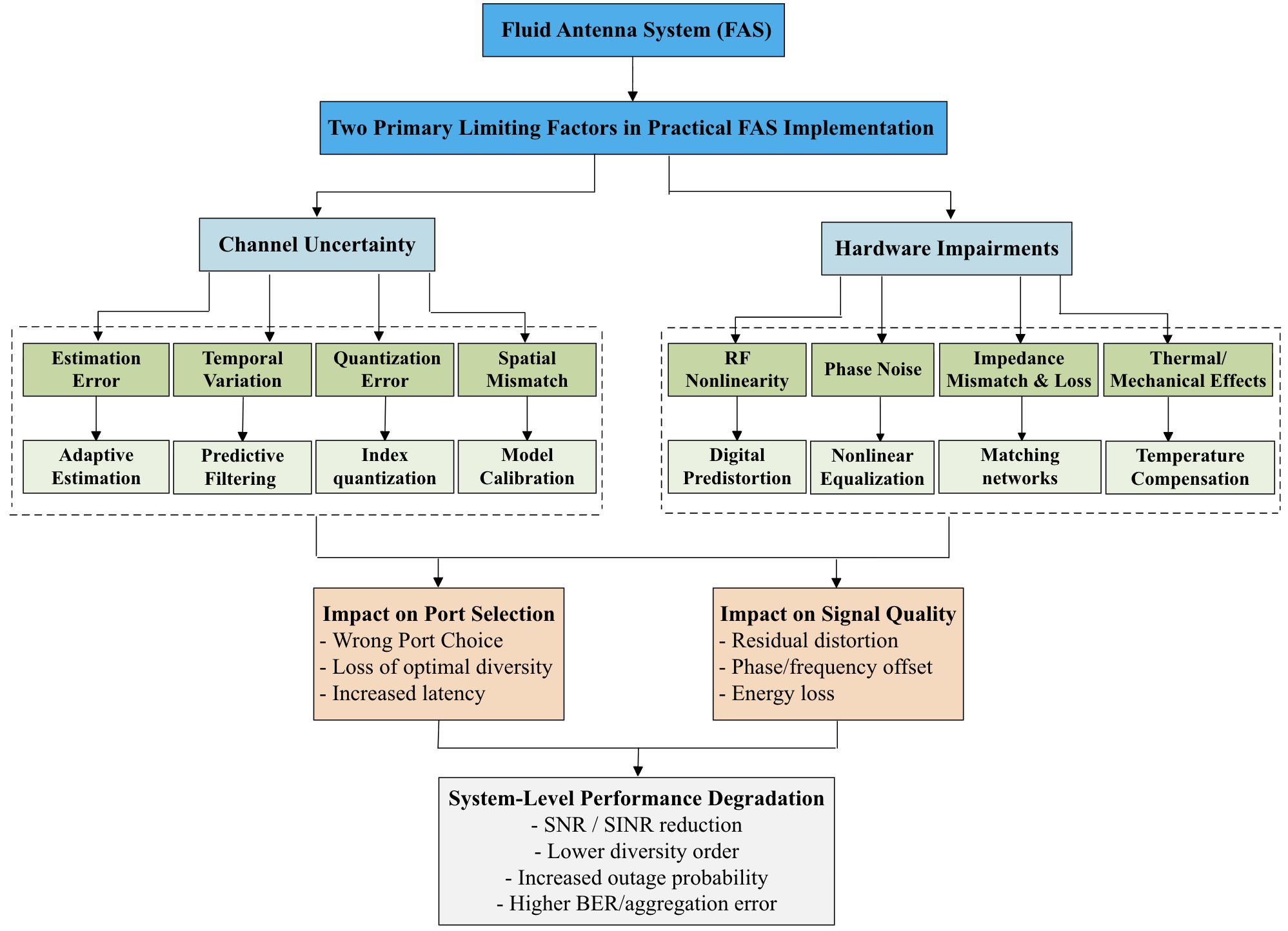} 
\caption{Comprehensive overview of channel uncertainty and hardware impairments in FAS and their impact on port selection and system performance.}
\label{fig:FAS_uncertainty_HW}
\end{figure*}

\subsection{Hardware Impairments}

\subsubsection{Sources of Hardware Impairments}

In FAS, the radiating structure dynamically changes its geometry through liquid metals, ionized solutions, or microfluidic channels. This physical reconfiguration introduces several hardware-level distortions:

\begin{itemize}
    \item \textbf{Nonlinearities in RF chains:} Power amplifiers, mixers, and low-noise amplifiers exhibit amplitude and phase distortion when operating close to saturation. These nonlinearities lead to intermodulation and spectral regrowth, especially when combined with the fluidic antenna’s tunable impedance behavior \cite{7106472, 11224643}.
    \item \textbf{Phase noise and frequency instability:} The mechanical motion of the conductive fluid can induce small but rapid variations in the electrical path length, resulting in phase jitter and frequency drift \cite{Bjornson2017MassiveMIMO}. Such instability directly impacts coherent detection and channel estimation accuracy.
    \item \textbf{Insertion loss and impedance mismatch:} The dynamic interface between liquid conductors and solid electrodes is rarely ideal. Viscosity, oxidation, or uneven surface tension may cause insertion loss or mismatch, reducing antenna efficiency and altering its radiation pattern \cite{8741198, 11247926, 11302793}.
    \item \textbf{Thermal and mechanical effects:} Temperature changes or repeated actuation cycles may modify the fluid’s conductivity, viscosity, or position accuracy, which in turn affects the reproducibility of the antenna’s performance \cite{11175437, 7297818, Mitchao2025ThermalAntenna}.
\end{itemize}

\subsubsection{Modeling Frameworks of Hardware Impairments}

To analyze and mitigate the impact of hardware imperfections, Several modeling approaches have been proposed to characterize hardware impairments:

\textbf{Aggregate Stochastic Models:}  
A widely adopted approach treats the combined effect of hardware imperfections as an additive distortion noise, often modeled as a zero-mean Gaussian process with variance proportional to the transmitted signal power. This representation captures the fact that stronger signals exacerbate nonlinear distortions and residual interference, enabling tractable analysis of achievable rate, SNR degradation, and error propagation in FAS \cite{Bjornson2017MassiveMIMO, 7106472}.

\textbf{Deterministic or Worst-Case Models:}  
In scenarios where safety margins or robust operation are critical, hardware imperfections can be bounded deterministically. For example, the distortion magnitude may be constrained within a known interval based on component specifications, allowing designers to optimize system performance under worst-case distortion conditions \cite{5447076, 11224643, 10753482}.

\textbf{Parametric Models of Specific Impairments:}  
For a more granular analysis, individual impairment types can be explicitly modeled. Common examples include polynomial models for power amplifier nonlinearities, phase-noise models for local oscillators, and error-vector-magnitude-based models for I/Q imbalance \cite{Zhu2025FluidAntennaGeometric, 11224643, 11196946}. Parametric models allow system designers to investigate the impact of specific hardware components on key FAS performance metrics such as port selection accuracy, spatial diversity, and aggregation error.

\subsubsection{Impact of Hardware Impairments on System Performance}

Hardware impairments directly influence the effectiveness of port selection and the overall FAS performance. Imperfections in the transmitter or receiver circuitry can introduce residual interference, nonlinear distortion, and biased channel estimates. Assuming ideal hardware yields an SNR of $\gamma_{\text{ideal}} = \max_{x \in X} \frac{P |h^{(t)}[x]|^2}{\sigma^2}$, whereas with hardware impairments, the realized signal-to-interference-plus-noise ratio (SINR) can be modeled as
$\gamma_{\text{impaired}} = \frac{P |h^{(t)}[x^{*}]|^2}{\sigma^2 + \eta P |h^{(t)}[x^{*}]|^2}$,
where $\eta$ quantifies the aggregate distortion factor and $x^{*}$ is the selected port. The corresponding SINR loss can be expressed as
$\Delta \gamma_{\text{HW}} = 10 \log_{10} \left( \frac{\gamma_{\text{impaired}}}{\gamma_{\text{ideal}}} \right)$,
highlighting the sensitivity of FAS performance to hardware quality.

Moreover, hardware impairments reduce the effective spatial diversity by limiting the accuracy of channel estimation and the precision of beamforming or over-the-air computation. Consequently, achievable rate, outage probability, and aggregation accuracy are all degraded, emphasizing the importance of jointly accounting for both hardware imperfections and channel uncertainty in system design.

\subsubsection{Strategies for Hardware Impairment Mitigation} 
Several techniques have been proposed to mitigate the influence of hardware imperfections in FAS:

\begin{itemize}
\item \textbf{Digital Predistortion and Nonlinear Equalization:}  
Nonlinearities introduced by power amplifiers, mixers, tunable components, or fluid-metal interfaces can cause amplitude/phase distortions and spectral regrowth \cite{10056864, Zhu2025FluidAntennaGeometric}. Digital predistortion and nonlinear equalizers compensate these imperfections by applying an inverse distortion profile to the transmitted signal. In the context of FAS, these techniques are particularly effective when the nonlinear characteristics vary with fluid displacement or channel geometry, ensuring consistent signal fidelity across different port positions \cite{10299674, 10677535}.

\item \textbf{Adaptive Impedance and Matching Calibration:}  
Continuous repositioning of the conductive fluid alters the local impedance profile along the waveguide or channels. This can degrade radiation efficiency, reduce SNR, and distort the desired spatial response. Real-time impedance tuning via programmable matching networks, varactor arrays, or reflection-coefficient sensing—allows the system to compensate for mismatch and maintain optimal coupling conditions \cite{Zhu2025FluidAntennaGeometric, 10309171, Hur2019AdaptiveImpedance, 9770295}. Such adaptive matching is crucial for FAS, where each fluid position corresponds to a unique electromagnetic boundary condition.

\item \textbf{Mechanical, Structural, and Thermal Stabilization:} 
Hardware degradation in FAS is strongly tied to environmental and mechanical factors: viscosity changes in the fluid, thermal expansion, channel deformation, and pressure fluctuations \cite{11224643, Tong2025FASHardware, AbuBakar2021LiquidAntenna}. Closed-loop actuation (using position sensors or microcontrollers), temperature compensation circuits, and materials engineered for thermal stability help preserve the fluid's shape, position accuracy, and electrical properties. These stabilization strategies reduce drift, prevent leakage or deformation under repeated actuation, and ensure consistent radiation characteristics over long-term operation.

\item \textbf{Joint Hardware–Algorithm Co-Design:}
Because hardware imperfections in FAS directly influence CSI quality, array response, and spatial diversity gain, algorithmic techniques must be designed with awareness of the underlying physical constraints. Co-design approaches, where beamforming algorithms, port-selection rules, over the air computation aggregation, or signal-scaling mechanisms explicitly incorporate hardware impairment models, enable robust performance under realistic conditions \cite{11281544, 10753482, Wang2025FluidAntenna}. Rather than treating impairments as external noise sources, such integration ensures that the communication strategy adapts to variations in actuator precision, fluid mobility, and matching accuracy.
\end{itemize}

The interrelated effects of channel uncertainty and hardware impairments discussed above are summarized in Fig.~\ref{fig:FAS_uncertainty_HW}, which provides a unified view of their origins, propagation paths, and their ultimate impact on the overall performance of FAS.

\section{Robust Optimization and Mitigation Strategies}

The performance degradation caused by channel uncertainty and hardware impairments in FAS underscores the need for robust and adaptive optimization frameworks. Such robust design approaches strive to ensure reliable performance under imperfect channel state information, nonlinear hardware distortions, and temporal or spatial variations inherent to the fluid antenna medium.

\subsection{Robust Optimization Frameworks}

Robust optimization frameworks in FAS are primarily categorized into worst-case and stochastic designs, both aiming to ensure reliable operation under channel uncertainty and hardware imperfections \cite{Pakravan2024OTAComputationFAS, 10474119, 10753482, Hu2025UplinkFASMIMO}. In the worst-case design paradigm, channel estimation errors are assumed to lie within a bounded uncertainty region, and the optimization problem is formulated to optimize performance metrics such as SNR or minimize MSE under the most adverse channel realization. For instance, if the actual channel vector is modeled as $\mathbf{h} = \hat{\mathbf{h}} + \Delta \mathbf{h}$, with $|\Delta \mathbf{h}| \leq \epsilon$, the robust beamforming solution guarantees stable performance for any admissible perturbation $\Delta \mathbf{h}$ within this set. This approach is particularly beneficial in FAS applications where correlation mismatches and unpredictable movements of the fluid antenna within its spatial container can lead to abrupt performance variations. Conversely, when statistical properties of the uncertainty are known, stochastic optimization or chance-constrained formulations are employed. The objective is to guarantee probabilistic reliability, such as maintaining a given SNR threshold with a probability exceeding $1-\delta$. This method captures realistic random channel fluctuations and mechanical noise in liquid metal movements. Together, these robust optimization strategies form the foundation for enhancing the resilience and adaptability of FAS under practical wireless conditions.

\subsection{Adaptive Port Positioning and Resource Allocation}

In dynamic wireless environments, adaptive control in FAS plays a vital role in mitigating the effects of channel uncertainty and hardware impairments. By jointly optimizing antenna port positioning, transmit power, and beamforming vectors, the system can continuously adapt to instantaneous channel variations and maintain reliable communication performance. Recent studies have demonstrated that gradient-based iterative algorithms can dynamically adjust port locations in response to real-time channel feedback, forming a closed-loop adaptive control mechanism \cite{Yan2025MovableAntenna, 11196915, 9715064, 10506795, 10753482}. Within this context, block coordinate descent methods are often employed to iteratively refine antenna positions and beamforming coefficients, while reinforcement learning techniques enable intelligent online decision-making under uncertain and non-stationary conditions. Moreover, model predictive control strategies can anticipate future channel states and fluid movement dynamics, allowing proactive adjustments of port locations and transmission parameters. Such adaptive methods can significantly improve robustness, especially under time-varying multipath channels and hardware-induced nonlinearities.

\subsection{Compensation for Nonlinear and RF Distortions}

To mitigate hardware impairments such as IQ imbalance, phase noise, and power amplifier nonlinearity, various compensation mechanisms can be incorporated into the control loop of FAS. Digital predistortion techniques are commonly employed to counteract amplifier nonlinearities prior to transmission, thereby preserving signal integrity \cite{7600411, 7297818}. Phase noise tracking methods leverage pilot-assisted estimators to correct oscillator drift and maintain coherent signal combining \cite{9503397}. Additionally, calibration-based approaches estimate and compensate for the combined effects of hardware mismatches and port coupling through either offline or online calibration procedures \cite{Tong2025FASHardware}. In FAS architectures, such self-calibration mechanisms are particularly crucial, as the dynamic motion of the conductive fluid can induce time-varying parasitic coupling among antenna ports \cite{11224643}. Integrating these compensation strategies within the FAS framework enhances overall system linearity, stability, and resilience to RF distortions.

\subsection{Robust Beamforming under Mobility and Delay}

FAS are inherently subject to response delays arising from liquid inertia and control signal latency, which may result in outdated CSI in rapidly time-varying environments. To address these challenges, robust beamforming strategies have been developed to enhance system adaptability and reliability under mobility-induced dynamics \cite{10021887, 10753082, Ahmadzadeh2025OTAFLMovableAntenna}. Predictive beamforming techniques exploit autoregressive channel models or machine learning–based predictors to forecast short-term CSI, thereby compensating for feedback and actuation delays \cite{11052222}. Hybrid analog–digital beamforming architectures further improve robustness by balancing computational efficiency with reconfiguration flexibility \cite{8371237, Wang2025FluidAntenna}. Moreover, error-tolerant signal aggregation methods particularly relevant in over-the-air computation and federated learning frameworks help reduce the sensitivity of system performance to delayed or imperfect updates \cite{Ahmadzadeh2025AIFASOTAFL, 11111711}. Collectively, these approaches enable FAS to maintain stable beamforming performance and efficient resource utilization in the presence of mobility, delay, and uncertainty.

\subsection{Integration with Machine Learning-Based Robust Design}

Machine learning (ML) and deep learning (DL) techniques have emerged as powerful enablers for robust optimization in FAS operating under channel uncertainty and hardware impairments \cite{Zhu2025FluidAntennaGeometric, 9264694, Ahmadzadeh2025AIFASOTAFL, 10299674}. By learning the complex, nonlinear relationships among channel states, antenna positions, and optimal transmission configurations, neural networks can effectively enhance system adaptability and decision-making in dynamic environments. Deep reinforcement learning enables autonomous port control and real-time resource adaptation under stochastic and non-stationary channel conditions \cite{ Ahmadzadeh2025OTAFLMovableAntenna, 11111711, saeid123456}. Graph neural networks provide a scalable means to capture spatial correlations and cooperative interactions among multiple fluid antennas in distributed FAS deployments \cite{9944643}. Moreover, physics-informed neural networks incorporate fluid dynamics and electromagnetic coupling principles directly into the learning process, thereby improving model interpretability and generalization \cite{Sun2020PhysicsConstrainedDL, Kochkov2021MLAcceleratedCFD}. Once trained, these data-driven frameworks offer fast inference, strong adaptability, and low-latency operation, positioning them as promising solutions for intelligent and real-time FAS control.



\section{Research Challenges and Future Directions}

Although FAS have shown remarkable potential for achieving dynamic spatial diversity and adaptability in future wireless networks, the presence of channel uncertainty, hardware nonidealities, and mechanical constraints still poses significant challenges for practical realization. This section outlines the open research issues and possible pathways for addressing them in forthcoming studies.

\subsection{Accurate Modeling of Spatio-Temporal Uncertainty}

A key challenge in FAS research lies in developing precise and tractable models that jointly capture the spatial mobility of the fluid antenna and the temporal variations of the wireless channel. Conventional stochastic channel models, which often assume static or quasi-static configurations, are inadequate for representing the continuous positional variability intrinsic to FAS. Future research should therefore focus on integrated electromagnetic–fluid dynamics models that describe the interplay between antenna motion, impedance variations, and channel response, as well as on non-stationary stochastic frameworks capable of capturing time-varying channel correlations influenced by fluid motion \cite{8371237, Pakravan2024OTAComputationFAS}. In addition, hybrid deterministic–statistical approaches that incorporate physical parameters such as actuation delay and viscous damping into stochastic channel representations are essential.

\subsection{Robust and Adaptive Control of Fluid Mobility}

The dynamic repositioning of fluid antennas introduces additional control and stability challenges, as antenna motion typically governed by electrocapillary, electrowetting, or mechanical pressure mechanisms is inherently influenced by inertia, viscosity, and thermal lag \cite{11175437, AbuBakar2021LiquidReconfigurable}. These factors can result in response delays and positioning inaccuracies, which in turn degrade link performance. Future research should investigate closed-loop control systems that leverage real-time feedback from both channel estimations and mechanical sensors to maintain precise positioning. Learning-based control frameworks, including reinforcement learning and adaptive model predictive control, can further enhance system adaptability by predicting and compensating for mobility-induced delays. Additionally, coordinating multiple antenna ports to dynamically balance electromagnetic gain and mechanical stability is critical, particularly under high-speed operation. Developing energy-efficient actuation mechanisms that maintain precise control without introducing excessive latency remains a key open problem for enabling practical and robust FAS deployment.

\subsection{Integration with Emerging Technologies}

The inherent adaptability and compactness of FAS make them well suited for integration with several emerging wireless technologies envisioned for 6G and beyond. In particular, FAS can complement RIS by combining continuous antenna mobility with large-scale reflection control to enhance coverage and enable precise beam manipulation \cite{9770295, 10794591}. In massive MIMO and cell-free networks, FAS-equipped nodes may serve as dynamic access points or edge devices, providing additional spatial diversity and flexibility \cite{9770295, 10146286}. For terahertz (THz) and millimeter-wave (mmWave) communications, the mobility of fluid antennas can help mitigate severe path loss and signal blockage, while in integrated sensing and communication scenarios, FAS motion can facilitate adaptive sensing and environmental awareness \cite{9264694, Shen2024MmWaveFluidAntenna}. Moreover, in over-the-air computation and federated learning frameworks, FAS can improve the robustness of gradient aggregation under channel uncertainty \cite{11111711, 11247926, 10474119}.

\section{Conclusion}

FAS represent a promising paradigm for future wireless networks, enabling dynamic spatial diversity and adaptive performance through continuous antenna reconfiguration. By exploiting spatial fluidity rather than fixed geometry, FAS can efficiently capture location-dependent channel variations, offering compact, energy-efficient, and flexible communication solutions for 6G and beyond. This review highlighted that the practical realization of FAS is strongly limited by channel uncertainty and hardware impairments. We examined advanced frameworks for modeling uncertainty, analyzed major hardware nonidealities such as nonlinearity and phase noise, and discussed robust design and optimization strategies to mitigate their combined effects. Building upon these insights, several open research directions have been identified. Future studies should focus on developing high-fidelity spatio-temporal channel models, integrating real-time control mechanisms for precise fluid manipulation, and advancing cross-domain co-design techniques that jointly optimize the mechanical, electromagnetic, and algorithmic layers.

\bibliographystyle{IEEEtran}
\bibliography{reffluid}

	\vfill
	
\end{document}